# Perovskite solar cells dominated by bimolecular recombination – how far is the radiative limit?

Kashimul Hossain[1], Dhyana Sivadas[2], Dinesh Kabra[1]* and Pradeep R. Nair[2]*

[1]Department of Physics, Indian Institute of Technology Bombay, Powai, Mumbai 400076, India.

[2]Department of Electrical Engineering, Indian Institute of Technology Bombay, Mumbai 400076. India.

**Abstract**:

Here, we report an experimental demonstration of perovskite solar cells dominated by bimolecular recombination and critically analyse their performance against radiative limits. To this end, we first establish a set of quantitative benchmark characteristics expected from solar cells limited by bimolecular recombination. Transient as well as steady state intensity dependent measurements indicate that our solar cells indeed operate at such limits with interface passivation comparable to the champion c-Si technology. Further, we identify novel characterization schemes which enable consistent back extraction of recombination parameters from transient optoelectrical and electroluminescence measurements. Remarkably, these parameters predict important features of dark current density *vs.* voltage characteristics $(J-V)$ and Suns-$V_{OC}$ measurements, thus validating the estimates and the methodology. Uniquely, this work provides a consistent and coherent interpretation of diverse experimental trends ranging from dark $J-V$, Suns-$V_{OC}$, steady state and transient intensity dependent measurements to electroluminescence quantum yield. As such, insights shared in this manuscript could have significant implications towards fundamental electronic processes in perovskite solar cells and further efficiency optimization towards Shockley-Queisser limits.

**Introduction:**

Among the various photovoltaic technologies, perovskite solar cells are the fastest-growing and have achieved efficiencies close to the market-leading c-Silicon solar cells ($\geq 25\%$) [1–3]. Excellent material and electronic properties of perovskites like high absorption coefficient, large diffusion length, low exciton binding energy, easily tuneable bandgap, etc.[4–6] and low cost fabrication[7–9] (solution, evaporation, spray deposition, etc.) have contributed to their broad appeal among photovoltaics community. The rate of progress and activity in this field is so impressive that performance close to the radiative or the Shockley-Quiesser (SQ) limit[10] could, possibly, be achieved in the near future. Indeed, solar cells limited by the radiative recombination are perceived as epitomes of perfection in photovoltaics. Practical solar cells involve multiple materials and fabrication steps with different thermal budgets. As a result, practical solar cell efficiencies are usually much lower than SQ limits – mainly due to the presence of defects in the bulk of the active material and imperfect interfaces.[10–12] Trap states or defects in a semiconductor [13,14] play a significant role in the solar cell efficiency as they directly contribute to non-radiative recombination mechanisms.[15,16]

A key feature of radiative recombination is its quadratic dependence on carrier density (under the assumption of $n = p$, where $n$ and $p$ are electron and hole densities, respectively). Hence, the order of the dominant recombination mechanism is a crucial information towards evaluating a solar cell against SQ limits. In this context, here we report the first experimental demonstration of perovskite solar cells dominated by bimolecular recombination under 1 Sun illumination. This experimental observation leads to several natural questions, such as (a) What criteria establish that a solar cell is indeed limited by bimolecular recombination (i.e., in addition to the unity light ideality factor)? (b) Is it possible to self-consistently back extract the relevant recombination parameters from device characteristics (i.e., not at a thin-film/spectroscopy level)? and (c) What might be the implications of these observations towards radiative limit performance? This manuscript addresses these issues through a combination of multi-probe characterization and theoretical analysis. Indeed, we provide a coherent interpretation of diverse characterization like dark $J - V$, Suns-$V_{OC}$, Intensity dependent open circuit voltage transients, and electroluminescence (EL) measurements. This analysis allows us to self-consistently back extract the recombination parameters from device level characteristics with interesting insights on further optimization.

Below, we provide the theoretical analysis of benchmark criteria to establish bimolecular recombination limited performance and its implications. To verify the same, we fabricated state-of-the-art perovskite solar cells with the active material $(FA_{0.83}MA_{0.17})_{0.95}Cs_{0.05}Pb(I_{0.83}Br_{0.17})$ of bandgap 1.6 eV. The details of materials and device fabrication are available in SI section A. Our devices report an efficiency of $\eta = 21.54\%$ under 1-Sun illumination with $J_{SC} = 23.35$ mA/cm$^2$, $V_{OC} = 1.160\ V$, $FF = 79.52\%$, over an active area of $0.175$ cm$^2$ (see **Figure S3a** and **Table S1**). The details of the photovoltaic (PV) performance, Incident Photon to Current conversion Efficiency (IPCE) spectrum, device active area *vs.* $V_{OC}$ & efficiency, Maximum Power Point Tracking (MPPT) stability, and PV parameters histogram are available in **Figure S3-S5** of SI (section B). Detailed analysis of the device characteristics under various scenarios and back extraction of recombination parameters from terminal characteristics are then discussed.

**Intensity dependent scaling laws**: Here, we consider optoelectrical characterizations in which the solar cell is maintained under open circuit conditions and is subjected to a pulsed illumination (**see Figure 1**). The carrier dynamics under illumination (with intensity $I$ and corresponding photo-generation rate $G$) for an undoped sample with $n = p$ is given as

$$\frac{\partial n}{\partial t} = G - k_1 n - k_2 n^2 - k_3 n^3 \qquad (1)$$

where $t$ is the time, and $n$ and $p$ denote the electron and hole density, respectively. The parameters $k_1$, $k_2$, and $k_3$ denote the monomolecular, bimolecular, and Auger recombination rates, respectively. While $k_1$ is contributed by mid-gap recombination centers, $k_2$ could be due to radiative and non-radiative recombination mechanisms. Auger recombination is expected to play a significant role at higher carrier levels and is considered later. Under such steady-state conditions ( and ignoring Auger recombination), with $G = G_0$, we have $G_0 = k_1 n_0 + k_2 n_0^2$ where $n_0$ is the steady-state carrier density. With $n = n_0 + \Delta n$, through a perturbation analysis[17–20] ($\Delta G \ll G_0, \Delta n \ll n_0$) we find

$$\frac{\partial \Delta n}{\partial t} = G' - (k_1 + 2k_2 n_0)\Delta n \qquad (2)$$

where $G' = \Delta G$ during $t < 0$ and $G' = 0$ during $t > 0$ (**see Figure 1**). During $t > 0$, the carrier transients are given by $\Delta n(t) = \Delta n(0)e^{-t/\tau}$ with $\tau = (k_1 + 2k_2 n_0)^{-1}$ and $\Delta n(0) \approx \Delta G T_{ON}$. As $V_{OC}$ is defined as the separation between quasi-Fermi levels, we have $V_{OC} = \left(\frac{kT}{q}\right) \times \ln\left(\frac{np}{n_i^2}\right)$, where $n_i$ is the intrinsic carrier concentration. With $n = p$, and under the assumption of small perturbation in comparison to the background illumination (see Section D of SI for detailed derivation) this leads to $\Delta V_{OC}(t) = \Delta V_{OC,max} e^{-t/\tau}$.

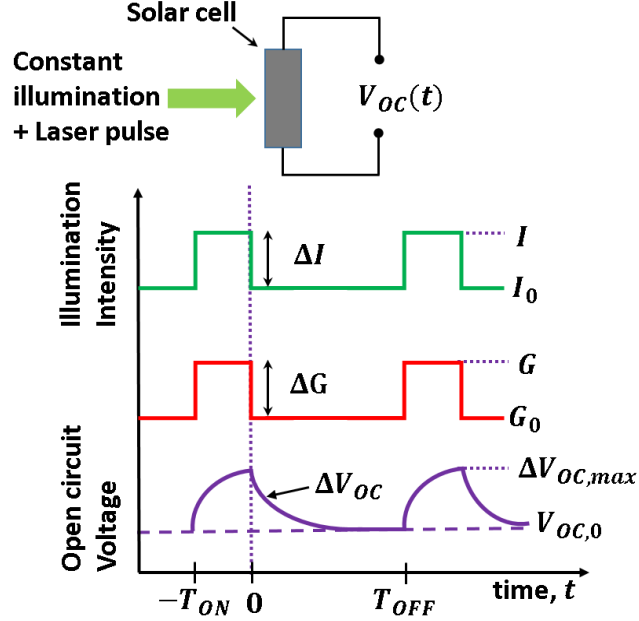

**Figure 1.** Opto-electronic characterizations to explore the dominant recombination mechanism. The top panel shows the schematic of the experimental setup (not drawn to scale). The solar cell is subjected to a constant illumination which is modulated by laser pulses and the resultant open circuit voltage transients are measured. The bottom panel shows the time dependence of illumination intensity ($I$), corresponding photo-generation rate ($G$) and open circuit voltage ($V_{OC}$).

Further mathematical analysis correlates the various parameters related to the steady state and transient measurements as (details are provided in Section D of SI. Note that the subscript 0 denotes steady-state parameters.)

$$V_{OC,0} = \frac{kT}{q}\ln(G_0) - \frac{kT}{q}\ln(k_2 n_i^2) \qquad (3a)$$

$$\Delta V_{OC,max} \approx 2\frac{kT}{q}\frac{\Delta G T_{ON}}{\sqrt{G_0/k_2}} \qquad (3b)$$

$$\tau^{-1} \approx k_1 + 2\sqrt{k_2}\sqrt{G_0} \qquad (3c)$$

$$\ln \tau \approx -\ln(2k_2 n_i) - \frac{q}{2kT}V_{OC,0} \qquad (3d)$$

With $G \propto I$, a perovskite solar cell with dominant bimolecular recombination should satisfy the following benchmarks in terms of the background illumination intensity $I_0$:

**(a)** light ideality factor is close to 1 (as per eq. 3a),

**(b)** $\Delta V_{OC,max}$ varies as $I_0^{-0.5}$ (as per eq. 3b),

**(c)** $\tau^{-1}$ varies as $I_0^{0.5}$ (as per eq. 3c), and

**(d)** the slope of $\ln(\tau)$ vs. $V_{OC,0}$ will be $\frac{q}{2kT}$ (as per eq. 3d).

**(e)** consistent back extraction of parameters like $k_1$, $k_2$, and $n_i$ from transient measurements. Further, eq. (1) indicates that monomolecular recombination dominates at low carrier densities. Accordingly, we expect that the device is dominated by monomolecular recombination under dark conditions and by bimolecular recombination under illumination. Such a device should exhibit the following additional characteristics.

**(f)** dark ideality factor is close to 2.

**(g)** ideality factor obtained using the Suns-$V_{OC}$ based pseudo-$J-V$ should be 1.

**(h)** The recombination parameters estimated using the illumination dependent steady state and transient measurements should anticipate/predict the reverse saturation current densities obtained from dark $J-V$ and the Suns-$V_{OC}$ based pseudo-$J-V$ characteristics. Further, we expect the above analysis to lead to a consistent explanation for the external quantum efficiency of electroluminescence of the same solar cell under dark conditions (i.e., when configured as a light emitting diode).

It is evident that the above list demands accurate and self-consistent analysis of multiple characterization techniques. This is clearly significant and relevant to the community as the above is expected from device level characterizations and not thin-film or material characterization. Hence, if successful, this could lead to self-consistent characterization of multiple phenomena based on terminal $J-V$ characteristics. To this end, below we provide a summary of the experimental results and then the required data analysis.

To check the experimental validation of theoretical predictions, we performed steady state as well as transient intensity dependent measurements, as described in **Figure 1**, on the perovskite solar cells having device structure ITO/Pz:PFN (9:1)/Perovskite/PCBM/BCP/Ag (**Figure S1**). Here, Pz:PFN (9:1) is the abbreviated form of mixed [4-(3,6-Dimethyl-9H-carbazol-9-yl)butyl]phosphonic acid (Me-4PACz) and poly(9,9-bis(3'-(N,N-dimethyl)-N-ethylammonium-propyl-2,7-fluorene)-alt-2,7-(9,9dioctylfluorene))dibromide (PFN-Br) in 9:1 volume ratio, which act as hole transport layer (HTL). Details regarding the fabrication process and device optimization are available in our earlier publication.[21] The photovoltaic device performance is provided in Section B of SI. The intensity dependent steady-state and transient studies are provided in **Figures S6**, and **S7** respectively (Section C of SI).

The results from device measurements are shown in **Figure 2**. Specifically, the variation of $V_{OC}$ with steady state illumination intensity is shown in **Figure 2a** while part (b) shows the dark $J-V$ characteristics (left y-axis) and pseudo-$J-V$ based on Suns-$V_{OC}$ (right y-axis)[22–26]. The key trends are as follows: (i) Light ideality factor is 1.05 (see **Figure 2a**) which compares well with the predictions of eq. 3a for devices dominated by bimolecular recombination[27]. (ii) **Figure 2b** indicates that the ideality factor associated with dark $J-V$ is close to 2 while the ideality factor associated with dark pseudo $J-V$ based on Suns-$V_{OC}$ measurements is close to 1. These observations clearly indicate that monomolecular recombination dominates the dark current while bimolecular recombination could dominate under illumination.

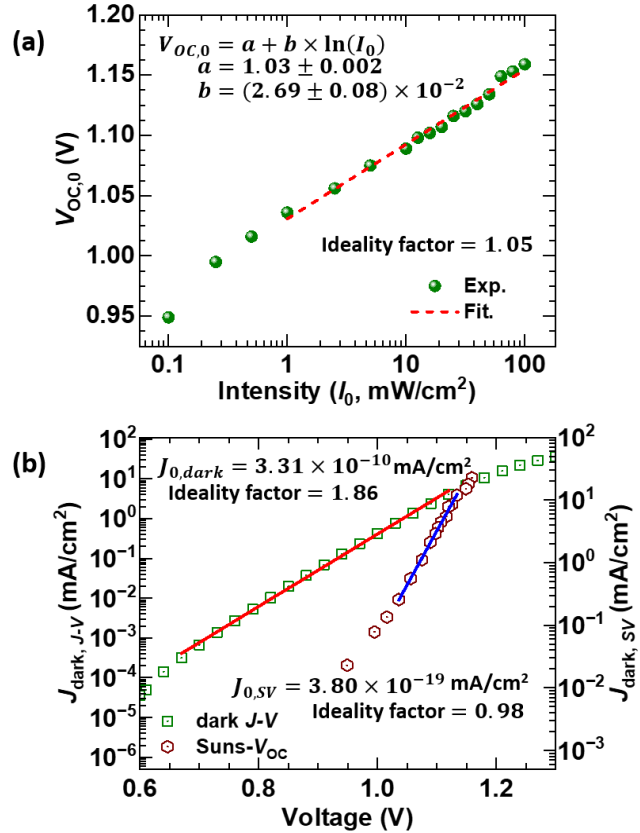

**Figure 2.** Intensity dependence of $V_{OC}$ and dark $J-V$ of perovskite solar cells. **(a)** Variation of the open-circuit voltage ($V_{OC,0}$) with steady state illumination intensity. The ideality factor obtained[27] from $V_{OC,0}$ vs. $\ln(I_0)$ is 1.05, which indicates that the device could be operating close to the radiative detailed balance limit. **(b)** The dark current characteristics measured from $J-V$ scans and Suns $V_{OC}$ measurements.[22–24] The $J_{dark,J-V}$ on the left y-axis indicates

measurement from dark $J-V$ scans whereas the $J_{dark,SV}$ on the right y-axis indicates the dark current estimated using Suns-$V_{OC}$ method.[25,26] Solid lines indicate numerical fits to obtain parameters like reverse saturation current density and ideality factor.

**Figure 3** summarizes the important trends from illumination intensity dependent transient measurements. Here, the open circuit voltage transients in response to a laser pulse are measured as a function of various background illumination intensities. Note that $\Delta I$ due to the laser pulse was kept a constant in our measurements (see Figure 1), as per literature.[28,29] Key parameters of the transients like $\Delta V_{OC,max}$ and $\tau$ are plotted in Figure 3 to contrast against the predictions of eq. 3 (see Figure S7 of SI for the transient data). The important features are (a) $\Delta V_{OC,max}$ varies as $I_0^\alpha$ (see **Figure 3a**, left panel) with $\alpha = -0.57$. The power exponent ($\alpha$) improves from $-0.57$ to $-0.54$ if the fit excludes the transients with $\Delta V_{OC,max} < 15\ mV$. This compares well with the theoretical predictions of eq. 3b. (b) $\tau^{-1}$ scales as $I_0^\beta$ (see **Figure 3a**, right panel) with $\beta = 0.48$. The power exponent ($\beta$) improves from $0.48$ to $0.51$ if the fit excludes the transients with $\Delta V_{OC,max} < 15\ mV$ - in accordance with the predictions of eq. 3c. (c) **Figure 3c** shows that $\tau$ varies exponentially with $V_{OC,0}$ with two distinct slopes. Indeed, for a broad range of measurements with $\Delta V_{OC,max} > 15\ mV$, we find that $\ln(\tau)$ varies linearly with $V_{OC,0}$ with a slope of $-19.94$ which compares well with $-q/2kT$, as per eq. 3d.

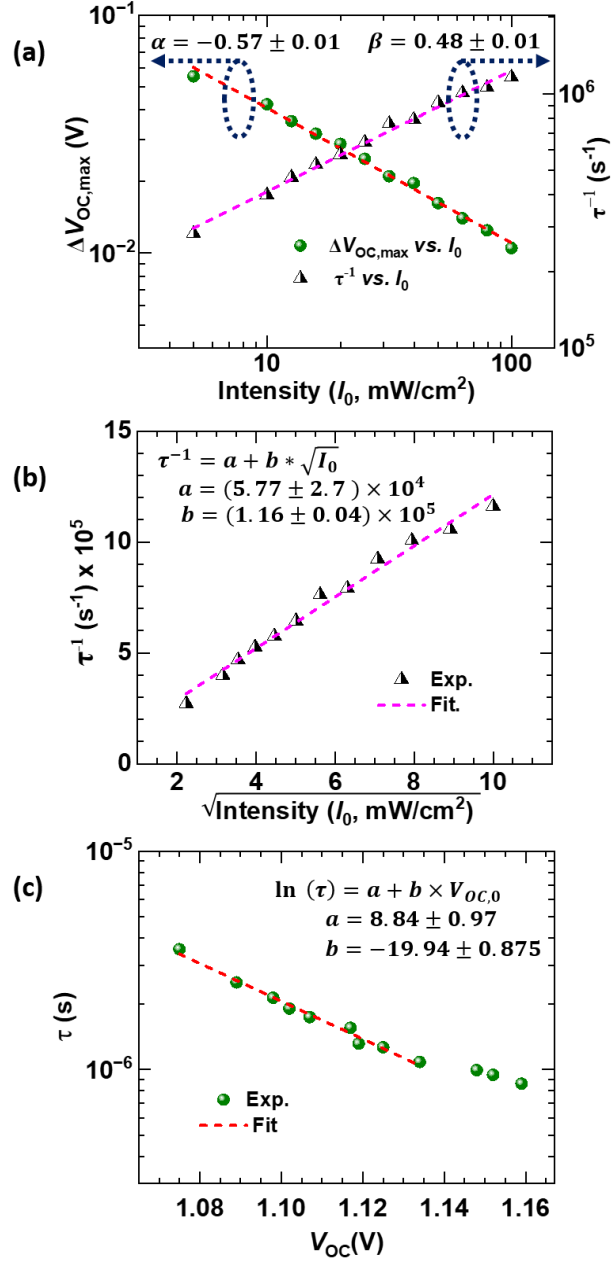

**Figure 3.** Comparison of experimental results against theoretical benchmark criteria (eq. 3) for perovskite solar cells. **(a)** $\Delta V_{OC,max}$ and $\tau^{-1}$ variation with $I_0$ are in accordance with eq. 3b,c. **(b)** Direct extraction of $k_1$ and $k_2$ using $\tau^{-1}$ vs. $\sqrt{I_0}$ plot, from the intercept and the slope, respectively (i.e., using eq. 3c). **(c)** $ln(\tau)$ varies linearly with $V_{OC,0}$ with slope close to $-q/2kT$, as anticipated by eq. 3d. These trends indicate that the device is limited by bimolecular recombination.

**Self-consistent back extraction of recombination parameters**: The results shown in **Figures 2-3** clearly indicate that our device is indeed limited by bimolecular recombination as the experimental trends compare well against all the benchmark criteria identified by eq. 3. In addition, eq. 3 also allows self-consistent back extraction of various recombination parameters (see Section E of SI for details). For example, as per eq. 3c, **Figure 3b** allows back extraction of $k_2$ as $(1.04 \pm 0.07) \times 10^{-10}$ cm$^3$s$^{-1}$ (from the slope) while the intercept leads to $k_1$ as $(5.77 \pm 2.7) \times 10^4$ s$^{-1}$ (see **Table S3**). Importantly, an independent estimate for $k_2$ is obtained using eq. 3b as $1.12 \times 10^{-10}$ cm$^3$s$^{-1}$ (**Figure S8**) which compares well with the previous estimate. This independent scheme for the estimation of $k_2$ is immensely useful for scenarios in which the direct estimation of $k_1$ and $k_2$ using eq. 3c could be limited by the noise in the measured data (such a case is explicitly discussed in **Figure S10-S16**, Section G of SI). From the intercept of **Figure 3c** and using eq. 3d, $n_i$ is estimated as $(0.25 - 1.98) \times 10^6$ cm$^{-3}$. Note that these estimates are obtained from transient measurements. Another estimate for $n_i$ can be obtained from steady-state measurements using eq. 3a and **Figure 2a** as the y-intercept of $V_{OC,0}$ vs. $ln(I_0)$ is related to $-ln(k_2 n_i^2)$. Using the back-extracted value for $k_2$ and the y-intercept of **Figure 2a** (See Table S3 for details), we find that $(1.03 - 1.19) \times 10^6$ cm$^{-3}$. Note that both these estimates for $n_i$ rely on the value of $k_2$. An independent estimate for $n_i$ can be obtained as follows: Specifically, $k_2 n_i^2$ can be obtained using eq. 3a and **Figure 2a** while $k_2 n_i$ can be obtained using eq. 3d and **Figure 3c**. This leads to an independent estimate for $n_i$ as $1.76 \times 10^6$ cm$^{-3}$. Interestingly and reassuringly, the estimate for $n_i$ obtained independently through respective steady state and transient measurements compare very well. Thus, through a combination of steady-state and transient illumination dependent measurements, here we obtain a coherent as well as consistent set of estimates for $k_1$, $k_2$, and $n_i$ (Section E and Table S3 for details).

The estimated recombination parameters are well in accordance with literature **Table S4**. It is worth noting that the estimates of $k_1$, $k_2$, and $n_i$ reported in the literature are typically based on the optical characterization of the thin films. Here, we estimated and validated the recombination parameters through a consistent set of device level experiments relying only on terminal characteristics, both steady state and transient. The parameters estimated from the steady-state and transient measurements compare well with each other and with the values reported in the literature (see Table S4).

As a further validation, the back extracted parameters can be used to predict the diode reverse saturation currents obtained from dark $J - V$ and Suns-$V_{OC}$ measurements. For this we use $k_1 = 5.77 \times 10^4$ s$^{-1}$, $k_2 = 10^{-10}$ cm$^3$s$^{-1}$, $n_i = 0.5 \times 10^6$ cm$^{-3}$, and $W = 450$ nm, where W is the thickness of the perovskite active layer. Note that the value of $n_i$ chosen is consistent with literature[30–32] and within the broad range estimated from our experiments. The ideality factor of dark current is close to 2 which indicates that the monomolecular recombination is dominant under such conditions (see **Figure 2b**). The reverse saturation current density measured from the dark $J-V$ is $3.31 \times 10^{-10}$ mA/cm$^2$, whereas the corresponding theoretical estimate for trap assisted recombination ($J_{0,SRH} = qk_1 n_i W$) is $2.08 \times 10^{-10}$ mA/cm$^2$. The ideality factor of dark pseudo $J-V$ characteristics is close to 1 which indicates that bimolecular recombination dominates under illuminated conditions **Figure S9a**. The corresponding reverse saturation current density obtained experimentally is $3.80 \times 10^{-19}$ mA/cm$^2$ which compares well with the theoretical estimate $J_{0,BB} = qk_2 n_i^2 W = 1.87 \times 10^{-19}$ mA/cm$^2$ (**see Figure S9b** and **Table S5**).[31]

With the measured value of the band gap ($E_g = 1.6\ eV$), (**Figure S2**) and $n_i \approx 0.5 \times 10^6$ cm$^{-3}$ with the assumption $N_C = N_V$, we find that $N_C \approx 1.65 \times 10^{19}$ cm$^{-3}$ were $N_C, N_V$ are the effective density of states of conduction and valence band respectively. It is well known that a solar cell remains limited by bimolecular recombination for all carrier densities $n_{BB} > k_1/k_2$. Using the back extracted values of $k_1$ and $k_2$, we find the corresponding value of $n_{BB} = 5.55 \times 10^{14}$ cm$^{-3}$. For 1 Sun illumination, the carrier density under open circuit conditions is given as $n_0 = \sqrt{G_0/k_2} = \sqrt{J_{SC}/qWk_2}$ ($W$ is the thickness of the active layer, $W = 450$ nm) which evaluates to $n_0 = 5.58 \times 10^{15}$ cm$^{-3}$. Hence, our device is dominated by bimolecular recombination for illumination intensities as low as 0.01 Sun. The carrier density at which Auger recombination dominates is given as $n_{Aug} > k_3/k_2$, where $k_3$ is the coefficient of Auger recombination[33,34]. For $k_3 = 10^{-28}$ cm$^6$s$^{-1}$, we find $n_{Aug} = 10^{18}$ cm$^{-3}$ which is much larger than $n_0$. This indicates that the relative contribution of Auger recombination under 1 Sun condition is negligible. All these factors contribute to ensure that our devices remain limited by bimolecular recombination over a large range of illumination intensities explored in **Figure 2.**

The SQ limit parameters[35,36] (i.e., with only radiative recombination), for a solar cell with a band gap $\sim 1.6\ eV$ and thickness $W \approx 500$ nm are $\eta \approx 30\%$, $J_{SC,SQ} \approx 26$ mA/cm$^2$, $V_{OC,SQ} \approx 1.3$ V, $FF_{SQ} \approx 90\%$. The recombination limited $J-V$ characteristics of a corresponding solar cell is given as

$$J_{AL} = -J_{SC,SQ} + qk_1 n_i W e^{qV/2kT} + qk_2 n_i^2 W e^{qV/kT} + qk_3 n_i^3 W e^{3qV/2kT} \quad (4)$$

Here, the first term on the RHS denotes the maximum $J_{SC}$ while the rest of the terms denote monomolecular, bimolecular, and Auger recombination, respectively (with the assumption that $n = p$ over the entire active layer). With the parameters $k_1 = 10^4$ s$^{-1}$, $k_2 = 10^{-10}$ cm$^3$s$^{-1}$, $k_3 = 10^{-28}$ cm$^6$s$^{-1}$, $n_i = 0.5 \times 10^6$ cm$^{-3}$, $W = 450$ nm, and $J_{SC,SQ} = 26$ mA/cm$^2$ the achievable limits (AL) of performance of such a cell is $\eta_{AL} = 27.76\%$, $V_{OC,AL} = 1.192\ V, FF_{AL} = 89.57\%$ (see Section H in the SI). With $k_1 = k_3 = 0$ (i.e., under dominant bimolecular recombination), the above estimate improves only marginally (**Figure S17** and **Table S8**). The gap between the achievable limit of efficiency with that of the SQ limit is of fundamental importance. Hence, exploration of the various electronic states and processes involved in the bimolecular recombination process is very relevant for identifying the achievable limit of efficiency and scope for further optimization.

In general, the bimolecular recombination rate could be due to radiative and non-radiative processes. Accordingly, we have

$$k_2 = k_{2,rad} + k_{2,nonrad} \qquad (5)$$

where $k_{2,rad}$ is the rate of band-band radiative recombination while $k_{2,nonrad}$ denotes non-radiative recombination processes[37–39] which vary as $n^2$. The origin of such non-radiative bimolecular processes is not well understood. Nevertheless, classical literature on trap assisted recombination indicates that shallow traps could lead to such effects[40]. Specifically, defects at an effective energy level $E_T$ could contribute to $k_2$ under the conditions $n_i e^{(E_T - E_i)/kT} > n_0$. With the back extracted value of $n_i = 0.5 \times 10^6$ cm$^{-3}$ and $n_0 = 5.58 \times 10^{15}$ cm$^{-3}$, we find that such traps should have $E_T - E_i > 0.59$ eV – rather the shallow traps could be with in 0.2 eV from the conduction/valence band edge for our active material with band gap of 1.6 eV. Preliminary transient photocurrent (TPC) measurements of our devices (**Figure S18**) indeed indicate the presence of shallow traps[41–43] in accordance with the above arguments. A device dominated by such shallow traps could result in a light ideality factor of 1. However, under dark conditions the expected ideality factor due to such shallow traps is 1. Experimental observation of dark ideality factor close to 2 (see Figure 2b) indicates that our devices have both mid-gap recombination centres and shallow traps. As a result, the dark ideality factor is 2 and the light ideality factor could be close to 1. Numerical simulation results provided in **Figure S19** of SI clearly support this inference.[44,45] Even though the hypothesis of shallow trap limited performance consistently explains the experimental observations on dark $J - V$, transient photovoltage, and transient photocurrent, it is evident that more experimental characterizations are needed to further establish and quantify role of shallow traps.

As non-radiative bimolecular recombination limits the performance of our solar cell, it is desirable to further characterize the parameters $k_{2,rad}$ and $k_{2,nonrad}$. This could be achieved through electroluminescence ($EL$) measurements[46] with the same devices being used as LEDs. Under forward bias conditions, the LED current could be dominated by recombination and is given by $J \approx qW(k_1 n + k_2 n^2 + k_3 n^3)$. The internal quantum efficiency of EL is given as

$$IQE_{EL} = \frac{k_{2,rad} n^2}{k_1 n + k_2 n^2 + k_3 n^3} \qquad (6)$$

The maximum value of $IQE_{EL}$, as per the above equation, occurs at $n_{EL} = \sqrt{k_1/k_3}$. With $k_1 = 10^5 \text{s}^{-1}$ and $k_3 = 10^{-28}$ cm$^6$s$^{-1}$, we have $n_{EL} \approx 3 \times 10^{16}$ cm$^{-3}$. At such carrier densities, it can be shown that bimolecular recombination with $k_2 = 10^{-10}$ cm$^3$s$^{-1}$ dominates both monomolecular and Auger recombination. Hence, further simple analysis indicates that the maximum external quantum efficiency of EL is given as

$$EQE_{EL,max} \approx f \frac{k_{2,rad}}{k_2} \qquad (7)$$

where $f$ is the outcoupling factor.[47,48] Eq. 7 allows direct estimation of $k_{2,rad}$ from the EQE$_{EL}$ measurements. The measured $EQE_{EL,max}$ as a function of the dark current of our devices (see **Figure S20** of SI) is 2.67%. With $f \approx 0.25$, we find $k_{2,rad} \approx 10^{-11}$ cm$^3$s$^{-1}$. With this estimate of $k_{2,rad}$, and using the equation $J = -J_{SC,SQ} + qk_{2,rad} n_i^2 W e^{qV/kT}$, the radiative limit performance parameters of our devices with back extracted parameters ($k_{2,rad} = 10^{-11}$ cm$^3$s$^{-1}$, $n_i = 0.5 \times 10^6$ cm$^{-3}$, $W = 450$ nm) are $\eta_{AL} = 29.34\%$, $V_{OC,AL} = 1.25\ V, FF_{AL} = 90.13\%$ (**Figure S21** and **Table S9**). These estimates are very close to the SQ limit performance for a solar cell of comparable bandgap. Interestingly, the above estimate is limited by uncertainties in the parameter combination $k_{2,rad} n_i^2 W$ which explains why the predicted $V_{OC}$ is lower than that of SQ limit of ~ 1.3 V (for the same bandgap). Further accurate estimates for $k_{2,rad} n_i^2 W$ could improve the $V_{OC}$ estimates thus approaching radiative detailed balance limits (i.e., $V_{OC,SQ} \approx 1.3\ V$).

**Discussions**

Our experimental results are well in accordance with the theoretical benchmarks identified and hence conclusively establish that our solar cells are dominated by bimolecular recombination. There are several novelties associated with this work.

It has been long established that the light ideality factor under steady-state conditions should be close to 1 for solar cells at dominated by bimolecular recombination.[49] Similarly, $ln(\tau)$ is also known to vary linearly[20] with $V_{OC,0}$ with a slope of $-q/2kT$. However, in isolation, these are not unique and sufficient conditions to claim device operation at radiative limit.[39] For instance, under low-level injection with an effective doping density $N_A$ and limited by trap assisted recombination, the $V_{OC,0}$ could vary as $kT/q \ln(G_0/k_1 n_i) + kT/q \ln(N_A/n_i)$ – which exhibits a light ideality factor of 1. However, as compared to eq. 3, it can be shown that such a solar cell would have distinctly different trends in transient characterizations and do not consistently anticipate the reverse saturation current densities estimated from dark $J - V$ and Suns-$V_{OC}$ methods. Hence, this work, for the first time, identifies a coherent set of benchmarks (i.e., not just a single criterion) expected from solar cells dominated by bimolecular recombination. These benchmark criteria are applicable for radiative limit performance as well. The low value for $k_1$ indicates that our devices have excellent carrier lifetime in the active layer and interfaces are well passivated. We remark that the light $J - V$ hysteresis is not significant in our devices (hysteresis index = 2.23%, Figure S3a, Table S1) - which also indicates good passivation of interfaces.[50] Under open circuit conditions with a negligible electric field, we have $k_1 = 1/\tau_B + 2S/W$, where $\tau_B$ is the trap assisted carrier lifetime in the active layer and $S$ denotes the recombination velocity at the interfaces with contact layers. Given $W = 450$ nm and the back extracted value of $k_1 = 5.77 \times 10^4$ s$^{-1}$, it is evident that the $\tau_B \geq 17$ μs and $S \leq 1$ cm/s. Such low values of $S$ are comparable to c-Si technology[51,52] while the $S$ reported[53,54] for perovskite thin films are of the order of $10^3$ cm/s.

The performance parameters of our solar cells under 1-Sun illumination are $\eta = 21.54\%$ with $J_{SC} = 23.35$ mA/cm², $V_{OC} = 1.160 V$, $FF = 79.52\%$, over an active area of 0.175 cm² and the bandgap of the active material used is 1.6 $e$V (see Figure S3a and Table S1). The performance of our device compares well with literature reports on solar cells with photoactive material of similar bandgap (see **Figure S3c** and **Table S2**). In comparison with the SQ or AL limits, it is evident that our devices differ in $J_{SC}$ and $FF$ due to reflection loss by the multilayer thin film devices and limited by the series resistance of the ITO coated glass substrates, respectively (see **Table S10**). The Suns $V_{OC}$ measurement offers a pseudo $J - V$ which is not affected by series resistance effects.[23,24] The FF estimated from the Suns $V_{OC}$ (89.02%) and achievable limit (89.57%) are similar **Figure S22a & Table S10**. The loss in $J_{SC}$ of the devices is due to reflection loss by the micro-cavity effect of the multi-layer thin film device **Figure S22b**.[55] Comparison with the achievable limits suggests areas of improvement. Evidently, there is much scope for improvement in the optical engineering of our solar cells (for example, the $J_{SC}$ can be further increased by using MgF$_2$ anti-refection coating towards the illumination glass side.)[56] which ensures optimum $J_{SC}$ for lower $W$ could be of interest. The FF gap indicates that resistive loss could be improved significantly.

Our devices are dominated by non-radiative bimolecular recombination. Hence, significant improvement in $V_{OC}$ is possible only if the shallow trap density is reduced. Indeed, these devices could result in near radiative limit performance if the shallow trap density can be reduced by one order of magnitude. This will result in $k_{2,rad} \approx k_{2,nonrad}$ and hence ~50% of the total recombination could be radiative, thus improving the efficiency further. Future research in this direction could be of broad interest.

Importantly, this work enables, for the first time, consistent estimation for $k_1, k_2,$ and $n_i$ from transient and steady state terminal $J-V$ characteristics of finished devices. To this end, we employed novel characterization schemes and theoretical analysis. Section G of SI shares results from a different device and the associated analysis which yields similar conclusions and parameters. Our estimates compare very well with the existing literature (Table S4).[57,58] Further, this manuscript identifies shallow traps as one possible cause of rather low $EQE_{EL}$ of otherwise good solar cells. Reduction in shallow trap density is indeed expected to improve the performance of both LEDs and Solar cells.

**Conclusions**

In summary, here we reported the first ever conclusive experimental demonstration of perovskite solar cells dominated by bimolecular recombination. This claim is supported by multiple experimental characterizations and detailed theoretical analysis. We proposed and validated coherent schemes for back extraction of recombination parameters from transient and steady state opto-electrical characterizations along with EL measurements which anticipate important parameters of dark $J-V$ and Suns-$V_{oc}$ measurements. Indeed, this work identifies achievable efficiency limits for perovskite solar cells with implications for both fundamental physics and process/device optimization.

**Acknowledgments:** This work was supported by the Ministry of New and Renewable Energy India: National Centre for Photovoltaic Research and Education (NCPRE). This work was also partially supported by the Indo-Swedish joint project funded by DST-India (DST/INT/SWD/VR/P-20/2019). PRN acknowledges the SERB project (CRG/2019/003163) and Visvesvaraya Young Faculty Fellowship.

**Supplementary Information:** SI contains the following information: Device fabrication and characterization protocols, photovoltaic device performance, details of steady-state and transient photovoltage measurements, a detailed derivation of eq. 3, back extraction of important parameters, analysis of the dark current, analysis of the achievable limits of the perovskite solar cell device performance, transient photocurrent and trap density, electroluminescence quantum efficiency etc.

**Conflict of Interest**

The authors declare no conflict of interest.

**Authors information**

**Corresponding Authors**

Prof. Pradeep R. Nair − Department of Electrical Engineering, Indian Institute of Technology Bombay, Mumbai 400076, India; orcid.org/0000-0001-9977-2737

Email: prnair@ee.iitb.ac.in

Prof. Dinesh Kabra − Department of Physics, Indian Institute of Technology Bombay, Mumbai 400076, India; orcid.org/0000-0001-5256-1465

Email: dkabra@iitb.ac.in

**Authors**

Kashimul Hossain − Department of Physics, Indian Institute of Technology Bombay, Mumbai 400076, India; orcid.org/0000-0002-4499-5308

Dhyana Sivadas − Department of Electrical Engineering, Indian Institute of Technology Bombay, Mumbai 400076, India;